\documentclass[twocolumn,showpacs,amsmath,amssymb,prd,aps]{revtex4}
\usepackage{graphicx}
\usepackage{epstopdf,color}

\newcommand{\req}[1]{Eq.\,(\ref{#1})}

\usepackage{comment}

\begin{document}

\title{Proposal for Resonant Detection of Relic Massive Neutrinos}

\author{Jeremiah Birrell}
\author{Johann Rafelski}
\affiliation{%
{Department of Physics, University of Arizona,  Tucson,  Arizona, 85721, USA}}
\date{\today}

\begin{abstract}
We present  a novel  method for detecting the relic neutrino background that takes advantage of  structured quantum degeneracy to amplify the drag force from neutrinos scattering off a detector.  Developing this idea, we present a characterization of the present day relic neutrino distribution in an arbitrary frame, including the influence of neutrino mass and neutrino reheating by $e^+e^-$ annihilation. We  present  explicitly the  neutrino velocity and de Broglie wavelength distributions for the case of an Earthbound observer.  Considering that relic neutrinos  could exhibit quantum liquid features at the present day temperature and density, we discuss the impact of neutrino fluid correlations on the possibility of resonant detection.
\end{abstract}
 \pacs{13.15.+g,95.85.Ry}

\maketitle

\section{Introduction}
Among the great science and technology challenges of this century is the development of the experimental capability to detect cosmic background neutrinos. To this end, we study the neutrino background as it is expected today, based on the freeze-out condition in the dense early Universe, combined with subsequent free-streaming and red-shifting to temperatures below the neutrino mass. In consideration of these results we argue that the present day massive relic neutrino background could form a quantum Fermi liquid state. We present arguments that the correlations inherent in a neutrino quantum Fermi liquid could be exploited in novel detection methods that utilize resonant amplification of relic neutrino drag forces.

This approach is in contrast to prior attempts to find a directly observable signature of the relic neutrinos, which have focused on the magnitude of the mechanical force due to scattering from a massless neutrino background~\cite{Opher,Lewis,Opher2,Stodolsky:1975,Cabibbo:1982,Shvartsman,Langacker:1983,Smith,Ferreras:1995wf,Hagmann:1999kf,Duda:2001hd,Gelmini,Ringwald:2009,Liao:2012,Hedman}.  The consensus reached in the literature is that, for an unstructured (relativistic) neutrino gas, such effects are far from the abilities of current detector technology.  Aside from elastic scattering there are also inelastic processes -- we note   the development of the PTOLEMY experiment~\cite{PTOLEMY} aiming  to observe relic electron neutrino capture by tritium, as originally proposed by Weinberg~\cite{Weinberg}. 

In this paper we will first characterize the free streaming distribution from the perspective of an observer in relative motion under the usual Boltzmann dilute gas assumption, utilizing the physically consistent equation of state from \cite{Birrell:2013_2}.   We will then argue that high degree of degeneracy of the non-equilibrium relic neutrino distribution, together with their temperature $T_\nu\ll m_\nu$, implies the inadequacy of the dilute gas assumption, resulting in a correlated background.  This leads us to explore the possibility of the detection of relic neutrinos by resonant amplification of the neutrino-detector interaction.

In our characterization of the cosmic neutrino background (CNB), we will focus on two important physical characteristics, namely the neutrino mass and effective number of  neutrinos $N_\nu$. $N_\nu\ne 3$ signals either a yet undiscovered massless particle inventory in the Universe, or as we will follow in this work, a transfer of  $e^+e^-$ annihilation entropy into neutrinos resulting in  their momentum distribution being `hotter' than generally accepted~\cite{Birrell:2013_2}. The magnitude of the  neutrino freeze-out temperature $T_k$ controls the amount of entropy that is transferred from $e^\pm$ into neutrinos before freeze-out. 
A numerical study based on the  Boltzmann equation  with two body standard model (SM) scattering~\cite{Mangano:2005cc} gives $N_{\nu}^{\rm th}=3.046$.  $N_\nu$ impacts Universe dynamics and recent  experimental  {\small PLANCK CMB} results contain several fits~\cite{Planck} which suggest that $N_\nu\simeq (3.30$--$3.62)\pm0.25$. {\small PLANCK CMB}  and lensing observations~\cite{Battye:2013xqa} lead to  $N_{\nu}=3.45\pm0.23$.

In consideration of the complexity of low energy neutrino interactions with the dense $e^\pm$ plasma near $T=O(1\text{MeV})$, we treated the  freeze-out temperature $T_k$ as a free parameter  and obtained  the dependence of the free streaming neutrino distribution on $T_k$ in~\cite{Birrell:2013_2}.  In particular, we find that the $T_k$ dependence can be expressed in terms of the measured  value  $N_\nu$ and vice verse. This was done under the assumption of a strictly SM-particle inventory and allowed for a determination of the effect of neutrino mass $m_\nu^i$, and  $N_\nu $ on the equations of state determining the Universe expansion. A value of $N_{\nu}\simeq 3.5$  can be interpreted in terms of a delayed neutrino freeze-out during the $e^\pm$ annihilation era. In the following we treat $N_\nu$  as a variable model parameter within the general observed experimental range and use the above mentioned relations to characterize our results in terms of $N_\nu$.

There are several available bounds on neutrino masses:\begin{itemize}
\item[a)] 
Neutrino energy and pressure components are important before photon freeze-out and thus $m_\nu$ impacts Universe dynamics. The analysis of cosmic microwave background (CMB) data alone leads to $\sum_i m_\nu^i<0.66$eV ($i=e,\mu,\tau$) and including Baryon Acoustic Oscillation gives $\sum m_\nu<0.23$eV~\cite{Planck}.  {\small PLANCK CMB} with lensing observations~\cite{Battye:2013xqa} lead to  $\sum m_{\nu}=0.32\pm0.081$ eV. 
\item[b)] 
Upper bounds have been placed on the electron neutrino mass in direct laboratory measurements  $m_{\bar\nu_e}<2.05$eV~\cite{PDG,Aseev}.\end{itemize}  
In the subsequent analysis we will focus on the neutrino mass range $0.05$eV to $2$eV in order to show that direct measurement sensitivity allows for the exploration of a wide mass range.

In section \ref{sec:dist_background} we introduce the free-streaming massive neutrino distribution.  We present our characterization of the neutrino distribution, including the dependence on $N_\nu$ and $m_\nu$, in section \ref{sec:dist}. In section \ref{sec:q_fluid} we show that at present day temperatures and densities, the CNB will have properties of a quantum Fermi fluid, including  density correlations. With this, we discuss the possibility of detection via resonance with the neutrino dynamical correlation frequency, which we estimate in the Earth frame to be within the range 110-160 MHz.

\section{The Free-Streaming Neutrino Distribution}\label{sec:dist_background}

The neutrino background and the CMB were in equilibrium until decoupling (called freeze-out) at $T_k\simeq { O}{\rm (MeV)}$. In the cosmological setting,  for $T<T_k$ the decoupled neutrino spectrum evolves according to the Fermi-Dirac-Einstein-Vlasov (FDEV) free-streaming  distribution~\cite{Langacker:1983,bruhat,Wong,Birrell:2013_2},
\begin{equation}\label{neutrino_dist_rest}
f(t,p)=\frac{1}{\Upsilon_\nu^{-1}e^{\sqrt{p^2/T_\nu^2+m_\nu^2 /T_k^2}}+ 1}.
\end{equation}
$\Upsilon_\nu$ is the  fugacity factor, here describing the underpopulation of neutrino phase space that was frozen into the neutrino FDEV distribution in the process of decoupling from the $e^\pm,\gamma$-QED background  plasma. The relation between $\Upsilon_\nu$ and $T_k$ is given in Ref.\cite{Birrell:2013_2}.  For $N_\nu\approx 3$, $\Upsilon_\nu$ is close to $1$, but for delayed freeze-out scenarios, $\Upsilon_\nu$ can deviate significantly form $1$.  Setting $T_k=0$ and $\Upsilon_\nu=1$ we obtain the baseline distribution used in~\cite{Ringwald:2004np}.

The neutrino effective temperature $T_\nu(t)= T_k\,(a(t_k)/a(t))$ is the scale-shifted freeze-out temperature $T_k$. Here $a(t)$ is the cosmological scale factor where $\dot a(t)/a(t)\equiv H$ is the observable Hubble parameter. 

The physical origin of the distribution \req{neutrino_dist_rest} is quite simple.  At the freeze-out time, when $T=T_k$, \req{neutrino_dist_rest} matches the equilibrium distribution.  After decoupling, each particle travels through the Universe on a geodesic, free of interactions other than gravity.  The effects of gravity are simply the usual red-shifting of momentum.  This implies that the distribution must be a function of $a(t)p$ only ($1/a(t)$ being the redshift temperature scaling) and so, given the initial condition at $T=T_k$, the necessary form is \req{neutrino_dist_rest}.

In practice, we will neglect $m_\nu/T_k\lesssim10^{-7}$.  This gives the distribution the same form as that of a massless fermion.  In other words, the fact that freeze-out occurred at $T_k\gg m_\nu$ means that the free-streaming distribution carries little information about the mass.  However, it is important to recall that we are not dealing with a textbook massless Fermi gas; all dynamical and kinematic observables, such as the energy density, pressure, velocity etc., will involve the neutrino mass.  This gives the results for a free-streaming particle a decidedly non-equilibrium nature, despite the familiar forms of some expressions.

An  important consequence of the structure of weak interactions is that $T_k\gg m_\nu$, meaning that freeze-out occurs prior to large scale annihilation of neutrinos.  This implies that the present day number density of neutrinos is  much higher compared to an equilibrium density.  This has important implications for the quantum nature of the CNB.   We will discuss this further in section~\ref{sec:q_fluid}.

\section{Characterizing the Neutrino Distribution in a Moving Frame}\label{sec:dist}
The neutrino background and the CMB were in equilibrium, mediated by $e^\pm$ plasma, until decoupling. Therefore one surmises that an observer would have the same relative velocity, $v_{\text{rel}}$, relative to the relic CNB  as with the CMB. Measurements of the CMB dipole anisotropy yield a relative solar system CMB velocity of $v_\oplus=369.0\pm 0.9$\,km/s \cite{Hinshaw}.  Taking into account the relative orientation of the earth orbital plane and the dipole direction (but neglecting ellipticity of the orbit), the relative velocity is modulated in the range $(-29.2,29.3)$\,km/s \cite{Bertone}.  In the following we will write velocities in units of $c$, though our specific results will be presented in km/s.

By casting the neutrino distribution, \req{neutrino_dist_rest}, in a relativistically invariant form we can make a transformation to the rest frame of an observer moving with relative velocity $v_{\text{rel}}$ to obtain
\begin{align}\label{neutrino_dist}
f(p^\mu)=&\frac{1}{\Upsilon_\nu^{-1} e^{\sqrt{(p^\mu U_\mu)^2-m_\nu^2}/T_\nu}+1}.
\end{align}
The 4-vector characterizing the rest frame of the neutrino FDEV distribution is
\begin{equation}\label{4_vel}
U^\mu=(\gamma,0,0,v_{\text{rel}}\gamma),\hspace{2mm} \gamma={1}/{\sqrt{1-v_{\text{rel}}^2}},
\end{equation} 
where we have chosen coordinates so that the relative motion is in the $z$-direction.   See~\cite{Safdi} and~\cite{Lisanti:2014pqa} for further discussion of the effects  of the neutrino distribution anisotropy in the Earth frame.

Using \req{neutrino_dist}, the normalized FDEV velocity distribution for an observer in relative motion is obtained has the form
\begin{align} \label{fvdistrib}
&f_v=\frac{g_\nu}{n_\nu 4\pi^2}\!\!\!\int_0^\pi \!\!\!\!\frac{ p^2dp/dv\sin(\phi) d\phi}{\Upsilon_\nu^{-1}e^{\sqrt{( E-v_{\text{rel}} p \cos(\phi))^2\gamma^2-m_\nu^2}/T_\nu}+1},\notag\\
&p(v)=\frac{m_\nu v}{\sqrt{1-v^2}},\qquad \frac{dp}{dv}=\frac{m_\nu}{(1-v^2)^{3/2}}.
\end{align}
where $g_\nu$ is the neutrino degeneracy and $n_\nu$ is the number density in the current frame.  $f_v$ was obtained by marginalizing over the angular variables, hence the integration over $\phi$.  The $\theta$ integral was performed analytically.

The normalization $n_\nu$ depends on $N_\nu$ but not on $m_\nu$ since decoupling occurred at $T_k\gg m_\nu$. For each neutrino flavor (all flavors are equilibrated by oscillations) we have, per neutrino or antineutrino and at non-relativistic relative velocity,
\begin{equation}\label{nnu}
n_\nu=[-0.36\delta N_\nu^2+6.7\delta N_\nu+56]\,{\rm cm}^{-3}
\end{equation}
($\delta N_\nu\equiv N_\nu-3$).  This is obtained by integrating the distribution \req{neutrino_dist} and using the relations between $\Upsilon_\nu$, $T_\nu$, and $N_\nu$ derived in Ref.~\cite{Birrell:2013_2}.

\begin{figure}
\centerline{\includegraphics[height=5.2cm]{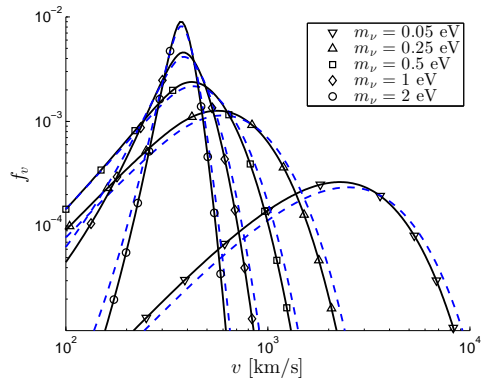}}
\caption{Normalized neutrino FDEV velocity distribution in the Earth frame. We show the distribution for $N_\nu=3.046$ (solid lines) and $N_\nu=3.62$ (dashed lines).}\label{fig:rel_v_dist_300}
 \end{figure}

We show $f_v$ in figure \ref{fig:rel_v_dist_300}   for several values of the neutrino mass, $v_{\text{rel}}=369$ km/s, and $N_\nu=3.046$ (solid lines) and $N_\nu=3.62$ (dashed lines).  As discussed above, the former corresponds to stardard model neutrino decoupling, while the latter corresponds to a delayed freeze-out scenario. From Ref.~\cite{Birrell:2013_2}, the respective fugacities are $\Upsilon_\nu=0.97$, $\Upsilon_\nu=0.71$. As expected, the lighter the neutrino, the more $f_v$  is weighted towards higher velocities with the velocity becoming visibly peaked about $v_{\text{rel}}$ for $m_\nu=2$ eV. 

\begin{figure}
\centerline{\includegraphics[height=5.2cm]{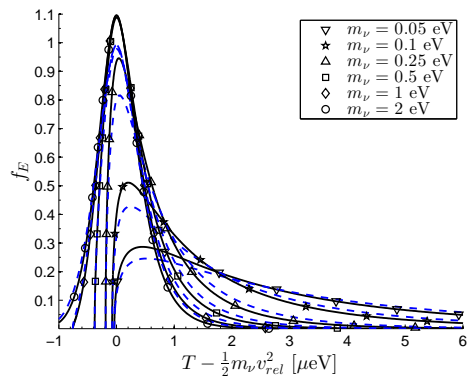}}
\caption{Neutrino FDEV energy distribution in the Earth frame. We show the distribution for $N_\nu=3.046$ (solid lines) and $N_\nu=3.62$ (dashed lines). }\label{fig:E_dist_300}
 \end{figure}

A similar procedure produces the normalized FDEV energy distribution $f_E$.  In \req{fvdistrib} we replace $dp/dv\to dp/dE$ where it is understood that 
\begin{equation}
p(E)=\sqrt{E^2-m_\nu^2},\qquad \frac{dp}{dE}=\frac{E}{p}.
\end{equation}
We show $f_E$ in figure \ref{fig:E_dist_300}  for several values of the neutrino mass, $v_{\text{rel}}=369$ km/s, and $N_\nu=3.046$ (solid lines) and $N_\nu=3.62$ (dashed lines). The width of the FDEV energy distribution is on the micro-eV scale and the kinetic energy $T=E-m_\nu$ is peaked about $T=\frac{1}{2}m_\nu v_{\text{rel}}^2$, implying that the relative velocity between the Earth and the CMB is the dominant factor for $m_\nu>0.1$ eV.

\begin{figure}
\centerline{\includegraphics[height=5.6cm]{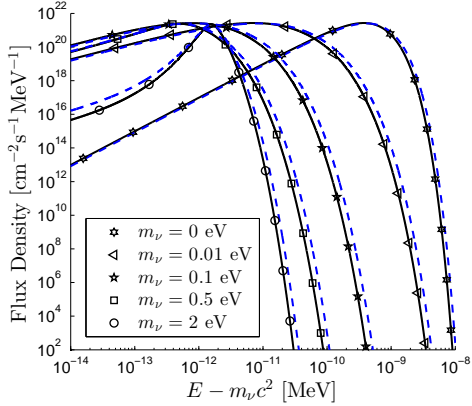}}
\caption{Neutrino flux density in the Earth frame. We show the result for $N_\nu=3.046$ (solid lines) and $N_\nu=3.62$ (dashed lines).}\label{fig:flux_dist}
 \end{figure}

By multiplying $f_E$ by the neutrino velocity and number density for a single neutrino flavor (without anti-neutrinos) we obtain the particle flux density,
 \begin{equation}
 \frac{dJ}{dE}\equiv\frac{dn}{dAdtdE},
\end{equation} 
shown in figure \ref{fig:flux_dist}. We show the result for $N_\nu=3.046$ (solid lines) and $N_\nu=3.62$ (dashed lines). The flux is normalized in these cases to a local density $56.6$~cm${}^{-3}$ and $60.4$~cm${}^{-3}$ respectively. The precise neutrino flux in the Earth frame is significant for efforts to detect relic neutrinos, such as the PTOLEMY experiment~\cite{PTOLEMY}. The energy dependence of the flux shows a large sensitivity to the mass. However, the maximal fluxes do not vary significantly with $m$. In fact the maximum values are independent of $m$ when $v_{\text{rel}}=0$, as follows from the fact that $v=p/E=dE/dp$.  In the Earth frame, where $0<v_\oplus\ll c$, this translates into only a small variation in the maximal flux.

\begin{figure}
\centerline{\includegraphics[height=5.1cm]{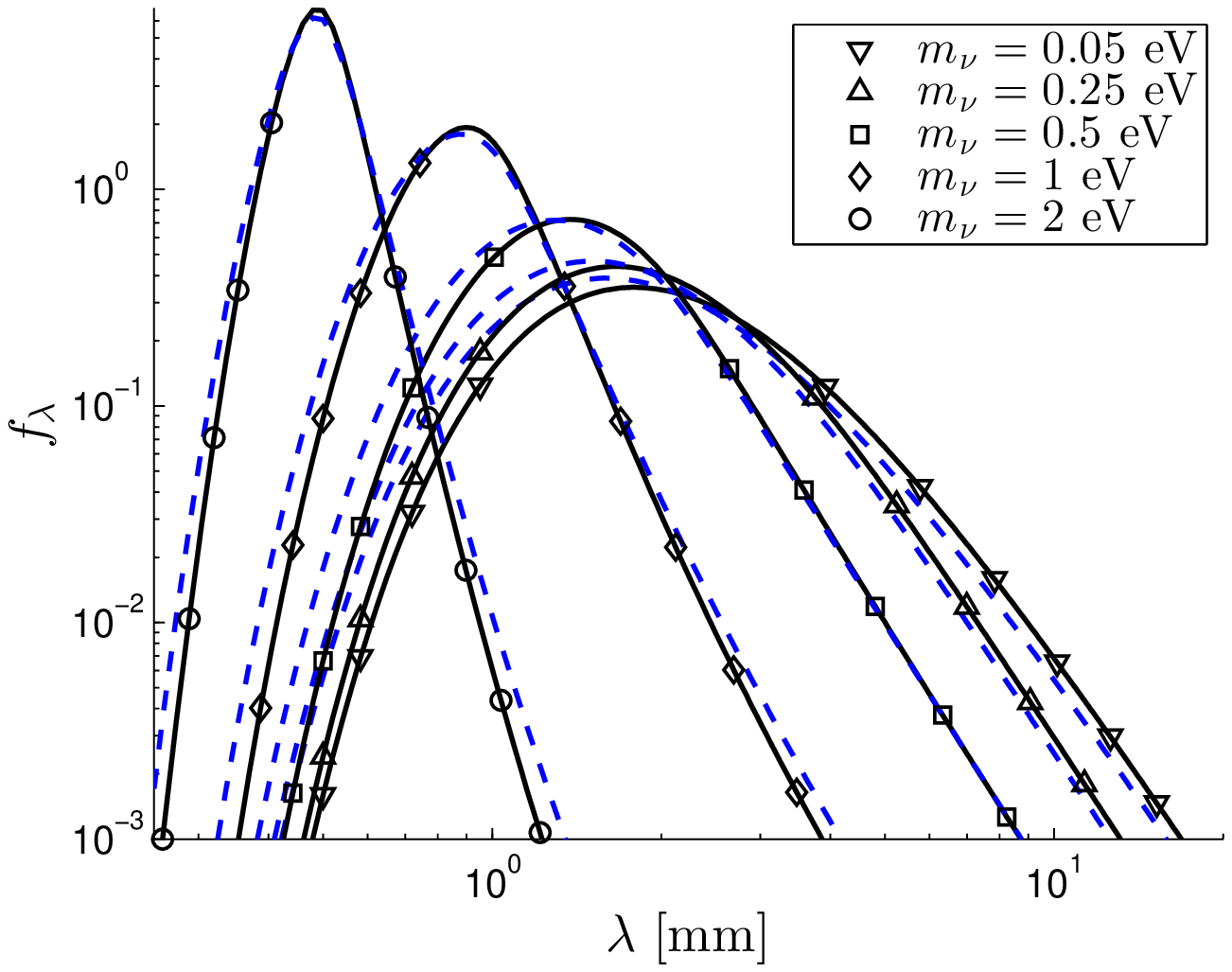}}
\centerline{\includegraphics[height=5.4cm]{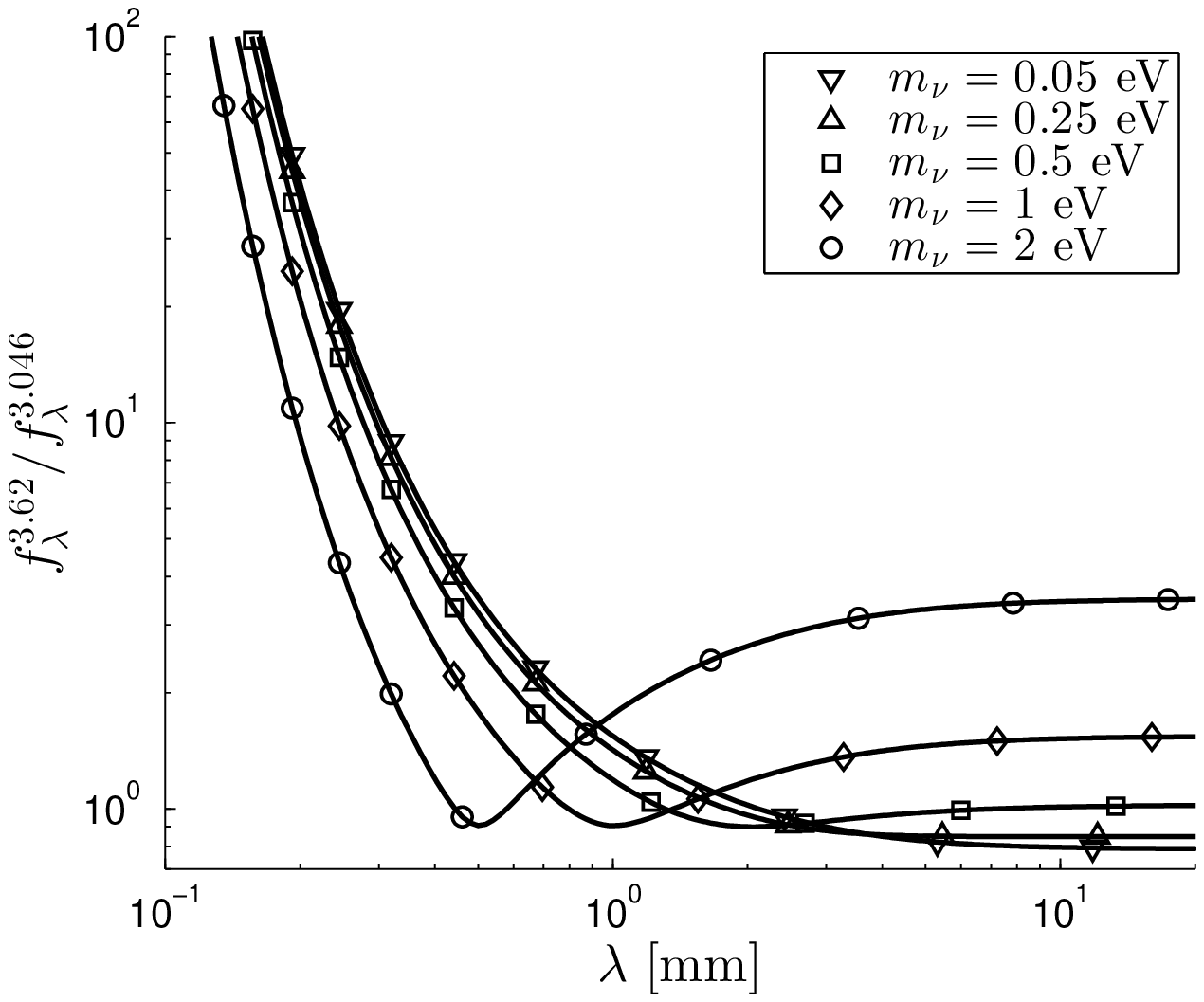}}
\caption{Neutrino  FDEV de Broglie wavelength  distribution in the Earth frame. We show in top panel the distribution for $N_\nu=3.046$ (solid lines) and $N_\nu=3.62$ (dashed lines) and in bottom panel their ratio.}\label{fig:deBroglie_300}
 \end{figure}

Using $\lambda=2\pi/p$ we find  in the the normalized FDEV de Broglie wavelength distribution
\begin{equation}
f_\lambda=\frac{ 2\pi g_\nu}{n_\nu\lambda^4}\!\!\int_0^\pi\!\!\! \!\frac{\sin(\phi) d\phi}{\Upsilon_\nu^{-1}e^{\sqrt{( E-v_{\text{rel}} p \cos(\phi))^2\gamma^2-m_\nu^2}/T_\nu}\!\!+\!1}
\end{equation}
shown in figure \ref{fig:deBroglie_300} for $v_{\text{rel}}=369$ km/s and for several values $m_\nu$ comparing  $N_\nu=3.046$ with $N_\nu=3.62$. 

\section{Neutrinos as a Quantum Fluid}\label{sec:q_fluid}


The prospects for detecting an unstructured relic neutrino background due to  neutrino scattering from a detector has been studied by many authors, both the $O(G_F)$ effects debated in  \cite{Opher,Lewis,Opher2,Cabibbo:1982,Langacker:1983,Smith,Ferreras:1995wf} and the $O(G_F^2)$ force in \cite{Shvartsman,Smith,Gelmini}, and were eventually found to be well beyond the reach of current detection efforts.

Considering the recent establishment of neutrino mass well above the scale of temperature of the neutrino background, a new physics opportunity arises which has not been considered in the earlier efforts.  While treating the scattering as a quantum process, all earlier studies modeled the CNB distribution as an unstructured particle gas.   As a consequence of this assumption, each neutrino impact on a detector is independent of the others. This means that irrespective of the spectrum structure we presented in the previous section, the impact to impact correlation of events in a neutrino detector  is a white shot noise spectrum. Hence only the overall strength of the interaction is relevant for neutrino detection. 

However, once one accepts that $m_\nu\gg T_\nu$, i.e. that neutrinos become `cold' in the free-streaming process while the Universe expands and cools,  one sees that the CNB neutrinos must form a dense quantum fluid. While we are first to note this in context of neutrino detection,  quantum CNB properties have already been recognized  in a different context~\cite{McElrath:2008ye}. 

\begin{figure}
\centerline{\includegraphics[height=5.6cm]{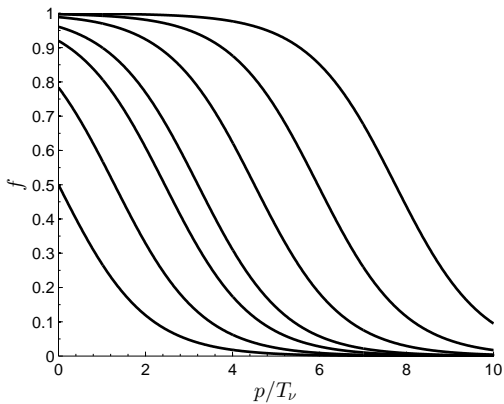}}
\caption{Neutrino distribution in the CMB frame with chemical potential corresponding to density enhancement by a factor of $1,2,5,10,25,50,100$ (bottom to top).}\label{fig:nu_dist_enhanced}
 \end{figure}

The expected degeneracy is further enhanced by gravitational accretion.  This has been studied previously for a classical particle  gas of neutrinos~\cite{Ringwald:2004np,Safdi}.  The former found density enhancement by a factor of $1-10$ for neutrino masses on the scale of $0.1$eV.   To quantify the importance of gravitational accretion  on the neutrino quantum fluid structure we show in figure~\ref{fig:nu_dist_enhanced} the neutrino distribution as a function of momentum normalized by temperature and corresponding to a density enhancement factor of 1, 2, 5, 10, 25, 50, 100 (bottom to top).  We see in figure~\ref{fig:nu_dist_enhanced}  that the quantum degeneracy increases rapidly and if indeed neutrinos accrete in the expected way, we are today immersed in a nearly degenerate quantum liquid. We return to this point further below.

A precise model of the background neutrino density correlations would require modeling their formation as the Universe cools while it expands and, equally importantly, incorporating the effects of solar system gravitational enhancement in the local quantum fluid distribution as discussed above. These steps are  beyond the scope of this initial proposal.  However, the presence of correlations in a quantum fluid is a well understood feature within the context of condensed matter physics~\cite{Kresse}, and we will base our arguments on this analogy --   we believe that the pivotal outcome of this circumstance can be captured by considering   neutrino impacts on the detector to be strongly correlated, leading to a structured impact  spectrum that can be leveraged for resonant detection. We will investigate this now within a simple model.

\begin{figure}
\centerline{\includegraphics[height=6.1cm]{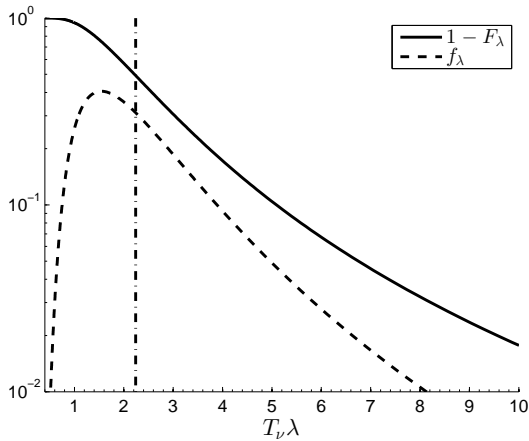}}
\caption{Neutrino de Broglie wavelength distribution (dashed line) and one minus the cumulative distribution (solid line), both computed in the CMB frame, with no density enhancement.  The vertical line shows $T_\nu/n_\nu^{1/3}$, a measure of the characteristic separation between neutrinos of a given flavor.  }\label{fig:deBroglie_num}
 \end{figure}

The importance of quantum effects is recognized in figure~\ref{fig:deBroglie_num} where we show the neutrino de Broglie wavelength distribution (dashed line) and cumulative distribution (solid line), both computed in the CNB rest frame for $N_\nu=3.046$.  The fact that we study now the neutrino liquid in the rest frame of the relic neutrinos where the distribution is a function of momentum only, see \req{neutrino_dist}, makes the result independent of neutrino mass.

The vertical line shows the value of $T_\nu L$ where the $L\equiv n_\nu^{-1/3}$, a measure of the average separation between neutrinos.  In computing $n_\nu$, the degeneracy  must be taken to be equal to one.  There is no spin degeneracy as   neutrinos (antineutrinos) are left (right) handed, meaning in this context that there is one neutrino state of each flavor.  (This is not the place to resolve the well known problem that handedness and helicity are not one and the same thing once neutrinos have a mass.)  We consider the density of only a single neutrino flavor at a time and without the corresponding antiparticle, so that we can ascertain the degree to which exclusion principle considerations make this a quantum system.  The conversion of one flavor into another (neutrino oscillation) does not affect this constraint  as for the present we assume that in equilibrium as many neutrinos oscillate out into another flavor as they oscillate in.

The cumulative distribution in figure~\ref{fig:deBroglie_num} shows that approximately $50\%$ of the neutrinos have deBroglie wavelength longer than $L$, indicating the importance of quantum effects in the homogeneous Universe, without any gravitational accretion. This result is independent of $T_\nu$ for $T_\nu<T_k$, as the redshifting of momentum and temperature compensate one another. In other words, the high density nature of the neutrino background in the early Universe, in terms of particles per deBroglie volume, is preserved in the subsequent free-streaming evolution because freeze-out occurs at $T_k\gg m_\nu$.  This important insight would be lost if one were to use (incorrectly) a thermal equilibrium neutrino distribution, since massive thermal neutrinos with $m_\nu>T_\nu$ are very dilute.
 
\begin{figure}
\centerline{\includegraphics[height=6.9cm]{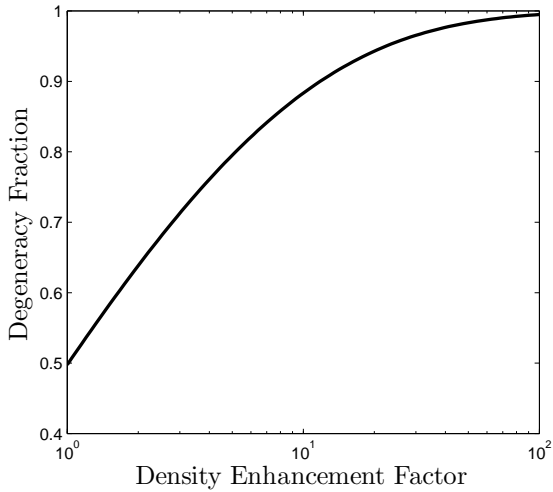}}
\caption{Fraction of neutrinos whose deBroglie wavelength is longer than the characteristic separation length, see figure \ref{fig:deBroglie_num},  as a function of the density enhancement factor compared to homogenous Universe density.}\label{fig:deBroglie_num2}
 \end{figure}

We now return to consider the effect of  density enhancement via gravitational accretion of neutrinos discussed above. Density increase (while maintaining the effective temperature) leads to a significant increase in degeneracy, and hence to strengthening of the quantum nature of the relic neutrino fluid.  We illustrate this in a quantitative fashion in figure~\ref{fig:deBroglie_num2}, where we show the fraction of neutrinos whose deBroglie wavelength is longer than the characteristic separation length   $T_\nu/n_\nu^{1/3}$, as a function of the density enhancement factor due to accretion. As already noted, on the left we see that about half of neutrinos are degenerate in a homogenous Universe and this fraction rises to 85\% for a factor of 10 neutrino density enhancement in the galaxy.

\section{Neutrinos Quantum Fluid Detection}\label{sec:q_fluid_detect}

We turn  now to evaluate the experimental opportunity provided by the newly recognized quantum correlations between neutrino impacts. A detailed  evaluation of the neutrino shot noise spectrum is beyond the scope of this work. However, much can be learned from a simple model.

First, we observe that neutrino oscillation measurements imply that at least one neutrino species has sufficiently high mass for the corresponding component of the CNB to be very cold, $m_\nu\gg T_\nu$~\cite{PDG}. The quantum nature of a cold massive fermion liquid will induce correlations in the CNB, even in the absence of interactions, due to  Pauli-Fermi exchange effect repulsion~\cite{feenberg2012theory} .  Such correlations will cause the temporal spectrum of neutrinos scattering from a detector to deviate from white noise.  We emphasize that we are not assuming that the potential (i.e. the detector) from which the neutrinos scatter is noisy, rather the noise is provided by the  stochastic process consisting of neutrino impacts on the detector.  For our discussion to be meaningful it is necessary that the detector is cooled to sufficiently low temperature so that the thermal noise in the detector does not wash out the signal from correlated neutrino impacts.

We make a crude estimate of the impact correlation  by assuming a perfect cubic lattice with side length $L=n_\nu^{-1/3}$.   In this case the neutrino impacts are not random and the characteristic frequency for such a lattice is 
\begin{equation}
f_{\text{shot}}= n_\nu^{1/3}v_{\text{rel}}
\end{equation}
which, as a function of $\delta N_\nu\equiv N_\nu-3$ and with $v_{\text{rel}}=369$km/s, is given by the following fit
\begin{equation}\label{shotnoise}
f_{\text{shot}}=(-0.43\delta N_\nu^2+5.6 \delta N_\nu+140)\frac{v_{\text{rel}}}{369 \text{km/s}}\text{MHz}.
\end{equation}
The annual Earth orbital velocity and daily rotation velocity produce a shifted frequency response as a function of $\delta v_{\text{rel}}\equiv v_{\text{rel}}-v_\oplus$ in the approximate range $\pm 30$km/s, and is an important signature that distinguishes a neutrino signal from other noise sources. This modulation, together with $0\leq \delta N_\nu\leq 0.8$, gives $129$MHz$\leq f_{\text{shot}}\leq 156$ MHz.  Gravitational accretion of neutrinos in the solar system would increase the frequency further, scaling with the cube root of the particle density.  Moreover, the increased quantum nature of the neutrino liquid further justifies the simple model we have used.

The central value estimate in \req{shotnoise} assumes that all neutrinos contribute to  the correlation effect. However, if only half of neutrinos participate in the correlated neutrino dynamics, as suggested by  figure \ref{fig:deBroglie_num}, then the central frequency would be reduced by a factor of $1/\sqrt[3]{2}$.  For $\delta N_\nu=0$ and $v_{\text{rel}}=369$km/s this corresponds to $111$ MHz.  On the other hand figure \ref{fig:deBroglie_num2} provides in quantitative fashion an opportunity to evaluate how the fraction varies with neutrino density enhancement. Thus measurement of resonance frequency comprises information about the neutrino quantum liquid.

The key point in our consideration is that such a structured neutrino background  noise raises the possibility of detection of the relic neutrinos via resonant amplification of the  momentum transfer from neutrino scattering.  We note that the peak at $f_{\text{shot}}=111$--$140$MHz for $\delta N_\nu=0$ and $v_{\text{rel}}=369$km/s is in a domain where one could argue an accidental noise signature could have already been observed.  This noise would arise from   neutrino elastic scattering resonating with electromagnetic mechanical resonators. Given that there is a strong cosmic background in this frequency range, see e.g. Ref.\cite{Furlanetto:2006jb},   an exploration of the background noise signature has probably not been undertaken\footnote{ Bruce G. Elmegreen, private communication}. We thus believe that in the interesting frequency range the variation of in-detector observed noise as a function of terrestrial observer velocity vector with respect to the CMB  is at present  unexplored.

To better understand the noise spectrum, including the very important resonance peak width, a significantly more precise model of the CNB correlations is needed.  We do not attempt to undertake this here.  However, we believe that the distinct frequency of the impact corelation   can vastly enhance the mechanical force when the effect is integrated over sufficiently long period of time and the detector has minimal damping.

\section{Discussion}
{\bf Remarks about dark matter:} While very different in detail, a challenge  similar to CNB detection exists with  the effort at direct detection of dark matter. Provided that some physical property of dark matter correlates the temporal mechanical impacts, the here proposed resonant mechanical force detection method also applies to  dark matter detection. 

Considering a dark matter particle  mass $M_D$, with lower limit of a few MeV~\cite{Boehm:2013jpa}, and the dark matter content in the Universe (20\% of all gravitating energy), one can estimate dark particle density. Assuming that  a sizable fraction of dark matter impacts is correlated we find  the impact frequency, $f_D$, is reduced to below MHz, and is decreasing as a function  of mass, $f_D\propto M_D^{-1/3}$. The dark matter and neutrino signatures are thus well seperated in frequency and both merit further experimental study by the here proposed novel resonant method.

{\bf Conclusions:} In this work we have characterized the relic cosmic  neutrino spectra in terms of their velocity, energy, and de Broglie wavelength distributions in a frame of reference moving relative to the neutrino background, with all examples focused on an Earth bound observer. We have shown explicitly the mass, $m_\nu$, dependence and the dependence on neutrino reheating expressed by $N_\nu$, choosing a range within the experimental constraints. 

The dependence on  $N_\nu$ and $m_\nu$ shown as a ratio on  linear scale in figure \ref{fig:deBroglie_300} (bottom frame) is sufficiently strong to suggest  that if and when relic neutrino detection becomes possible both  $N_\nu$ and $m_\nu$ would be directly measurable. The effect of $N_\nu$ as presented here is to increase neutrino flux~\cite{Birrell:2013_2}, see \req{nnu}. However, to this end one must gain precise control over the enhancement of neutrino galactic relic density due to  gravitational effects~\cite{Ringwald:2004np} as well as the annual modulation~\cite{Safdi}.  

As we argued in section \ref{sec:q_fluid}, the  relic  background of massive neutrinos will not behave as a purely free streaming gas, but have properties of a quantum liquid.  This inevitably leads to  impact correlations which will cause the power spectrum due to neutrinos  scattering off a detector to deviate from white noise, as we discuss in section \ref{sec:q_fluid_detect}. The spectrum should exhibit a peak near a frequency  arising from the relative velocity of the terrestrial observer  with respect to CNB. Based on a simple model we find an expected range of  111 -- 140MHz for  the central value, with the frequency increasing based on gravitational accretion enhanced neutrino density. The frequency is subject to modulation by the temporal variation of the observer velocity vector with respect to CNB.   Such correlated structure in the noise spectrum opens the possibility for resonance based CNB detection methods, amplifying the otherwise negligible effect of neutrino drag.


\begin{acknowledgements}
This work has been supported by a grant from the U.S. Department of Energy, DE-FG02-04ER41318
and was conducted with Government support under and awarded by DoD, Air Force Office of Scientific Research, National Defense Science and Engineering Graduate (NDSEG) Fellowship, 32 CFR 168a.
\end{acknowledgements}



\end{document}